Title: Changes to the TN Middle Grades Science Curriculum According to NGSS

Larry L. Bowman, Jr.
East Tennessee State University, NSF GK-12 STEM Teaching Fellow
And
Aimee Lee Govett, Professor
Center for Excellence in Mathematics and Science Education
East Tennessee State University

*Introduction*

The National Resource Council (NRC) of the National Academy of Sciences developed the Framework for K-12 Science Education (July 2011) as the first step in the creation of the Next Generation Science Standards (NGSS). The Framework draws on current scientific research and the ways students learn science effectively.  NGSS, released in Spring 2013, demonstrate a substantial consensus on what all students should know and be able to do at specific K–12 grade levels and what topics can be postponed or excluded on the premise of developmental readiness.  Tennessee was one of the 26 lead state partners to provide leadership to the standards writing team and to provide guidance as states deliberate on the adoption and implementation of the NGSS.

In order to facilitate this work, one of the STEM Fellows, Larry Bowman (author), created a series of posters showing correlations between the current Tennessee Science Education standards (Huffman, 2009) and NGSS (Achieve, 2013). These 18 posters serve as a series of guidemaps between the Tennessee Curriculum Standards for Science Education (TNCSSE) for grade levels kindergarten through high school and the corresponding NGSS. This article specifically addresses the NGSS Middle Grades Disciplinary Core Ideas (DCIs) as compared to the existing 6-8 Science Standards from the Tennessee Science Curriculum

Framework. Some Grade Level Expectations (GLEs) for Middle Grades Science have been relocated to other grade levels, and others have been introduced. We will examine the underlying principle and research based justification behind this.

DCIs are grouped in four domains: the physical sciences; the life sciences; the earth and space sciences; and engineering, technology and applications of science. Middle Grades curriculum and the research surrounding it are somewhat of a conundrum. Research related to teaching science in elementary grades covers students ranging from 5 to 12 years of age or grades 1 to 6 or 7 in most educational systems. The term Middle Grades is used for students 10 to 15 years of age or grades 5 or 6 through 8. Thus, there is some overlap. We will address the upper elementary years because this grade band covers the critical years when children are transitioning from concrete operational to formal operational stages. Preadolescent children demonstrate logical, concrete reasoning in the concrete operational stage. Children also develop operational thinking, the ability to perform reversible mental actions (Piaget and Inhelder 2013).

During this stage, however, most children still cannot tackle a problem with several variables in a systematic way. In the formal operational stage, adolescents are able to logically use symbols related to abstract concepts, such as algebra and science. They can think about multiple variables in systematic ways, formulate hypotheses, and consider possibilities (Piaget and Inhelder, 2013). While concrete concepts can be demonstrated with an object, a picture, or by pantomiming, abstract concepts exist in thought alone and are more challenging to teach (Archer and Hughes, 2011). Piaget and Inhelder determined that children in the concrete operational stage were fairly good at the use of inductive logic (2013). Inductive logic involves going from a specific experience to a general principle. Children in the formal operational stages begin using deductive logic, which involves using a general principle to determine the outcome of a specific event (Piaget and Inhelder, 2013).

*Changes in Curriculum Standards*

The Tennessee Science Standards involving physical properties of minerals and elements, compounds, and mixtures are covered in the NGSS DCIs for fifth grade physical science (Table 1). This reflects the cognitive theory that early adolescent minds learn best through concrete examples and experiential learning. Physical properties of minerals can be investigated in a very concrete way.

**Table 1:** Middle Grades Science curriculum relocations with the adoption of Next Generation Science Standards (NGSS). Grade Level Expectations (GLEs) and the corresponding relocated NGSS Disciplinary Core Ideas (DCIs) in 5th grade courses (5).

| Tennessee Standards Relocated | | New Location within NGSS | |
|---|---|---|---|
| *Earth and Space Sciences* | | | |
| **GLE 0707.7.1** | Describe the physical properties of minerals. | **5-PS1-3** | Make observations and measurements to identify materials based on their properties. |
| *Physical Sciences* | | | |
| **GLE 0807.9.4** | Distinguish among elements, compounds, and mixtures. | **5-PS1-3** | Make observations and measurements to identify materials based on their properties. |
| | | **5-PS1-4** | Conduct an investigation to determine whether the mixing of two or more substances results in new substances. |

With the adoption of NGSS, seven current middle grades standards will be moved to the secondary level (Table 2). Many of these concepts are abstract in nature and are moved to subject specific content courses, such as Biology and Physics courses. All standards that are relocated do not lose content but are instead moved to more appropriate grade-level. The standards in this category are a reshuffling of difficult abstract concepts to the high school level where students are more capable and cognitively able to digest the rationale behind why science and nature work in the ways they do. In middle grades memorization of abstract material is possible but its digestion, and thereby learning, is lessened (National Research Council, 2012).

**Table 2:** Middle Grades Science curriculum relocations with the adoption of Next Generation Science Standards (NGSS). Grade Level Expectations (GLEs) and the corresponding relocated NGSS Disciplinary Core Ideas (DCIs) to High School courses (HS).

| Tennessee Standards Relocated | | New Location within NGSS | |
|---|---|---|---|
| *Life Sciences* | | | |
| GLE 0707.1.4 | Illustrate how cell division occurs in sequential stages to maintain the chromosome number of a species. | HS-LS1-4 | Use a model to illustrate the role of cellular division (mitosis) and differentiation in producing and maintaining complex organisms. |
| *Physical Sciences* | | | |
| GLE 0607.12.1 | Describe how simple circuits are associated with the transfer of electrical energy. | HS-PS3-1 | Create a computational model to calculate the change in the energy of one component in a system when the change in energy of the other component(s) and energy flows in and out of the system. |
| GLE 0601.12.2 | Explain how simple electrical circuits can be used to determine which materials conduct electricity. | HS-PS3-2 | Develop and use models to illustrate that energy at the macroscopic scale can be accounted for as a combination of energy associated with the motions of particles (objects) and energy associated with the relative positions of particles (objects). |
| GLE 0707.11.1 | Identify six types of simple machines. | HS-ETS1-2 | Design a solution to a complex real-world problem by breaking it down into smaller, more manageable problems that can be solved through engineering. |
| GLE 0807.9.5 | Apply the chemical properties of the atmosphere to illustrate a mixture of gases. | HS-PS1-3 | Plan and conduct an investigation to gather evidence to compare the structure of substances at the bulk scale to infer the strengths of electrical forces between particles. |
| GLE 0807.9.6 | Use the periodic table to determine the characteristics of an element. | HS-PS1-1 | Use the periodic table as a model to predict the relative properties of elements based on the patterns of electrons in the outermost energy level of atoms. |
| GLE 0807.9.9 | Explain the basic difference between acids and bases. | HS-PS1-1 | Use the periodic table as a model to predict the relative properties of elements based on the patterns of electrons in the outermost energy level of atoms. |

For example, only one *Life Sciences* standard is moved from middle grades to secondary: GLE 0707.1.4 (Table 2). This standard addresses the understanding that cell division occurs in specific stages in order to maintain chromosome number. The current standard and the NGSS standard read very similarly, the former asking for an illustration and the latter asking for a model. However, the change with the adoption of NGSS would be that high school freshman are capable of addressing the intricacies and mathematics involved in the sequential halving and doubling of chromosomes in addition to the replication steps involved in cell division. Whereas, a 7th grader may have difficultly mastering the steps in cell division alone and would likely not understand the implications of a replication regime designed to maintain chromosome number aside from memorization of steps.

Six of the current standards in *Physical Sciences* have been moved to the secondary level and are designed to be included in a Physical Science, Chemistry, or Introductory Physics class. These topics include circuitry, simple machines, gas mixtures, the periodic table, and acids and bases. Again, these concepts represent abstract properties of the natural world that are not immediately accessible to middle grade students but are to high school students. The concepts behind the physical properties of matter are most abstract and without the subsequent steps of their application involved in the learning process, their importance is diminished to rote memorization in the middle grades. Moving these standards to content-focused classes such as Chemistry and Physics allows for students to rationalize why these properties exist and why they are important in the natural world, as opposed to being merely able to learn facts without digesting them (National Research Council, 2012).

Additionally, as NGSS moves these standards to the high school grade level, we see a more engaged trajectory for the students, delving into modeling and conducting experiments to completely understand the content knowledge and apply it to everyday problems. This allocation of time to learning the rationale behind the core sciences is an imperative part of keeping students engaged in the sciences and to foster natural inquiry that will be important for future career goals. We hope that the scientific and engineering practices will be further emphasized with the adoption of NGSS.

Some standards would be new additions to the Tennessee curriculum entirely (Table 3). Earth and Space Sciences and Physical Sciences would incorporate new standards that address modern technologies and environmental issues. The earth science topics mentioned in the additions include earth's geologic history and issues of environmental concern such as how natural hazards will and do affect society and how earth's resources are unevenly distributed. These DCIs incorporate important policy issues and resource allocation issues that will face individuals with future careers in the sciences.

**Table 3:** Middle Grades Science curriculum additions with the adoption of Next Generation Science Standards (NGSS) by discipline.

| | **Additions** |
|---|---|
| | *Life Sciences* |
| **None** | |
| | *Earth and Space Sciences* |
| **MS-ESS1-4** | Construct a scientific explanation based on evidence from rock strata for how the geologic time scale is used to organize Earth's 4.6-billion-year-old history |
| **MS-ESS3-1** | Construct a scientific explanation based on evidence for how the uneven distributions of Earth's mineral, energy, and groundwater resources are the result of past and current geosciences processes. |
| **MS-ESS3-2** | Analyze and interpret data on natural hazards to forecast future catastrophic events and inform the development of technologies to mitigate their effects. |
| | *Physical Sciences* |
| **MS-PS1-3** | Gather and make sense of information to describe that synthetic materials come from natural resources and impact society. |
| **MS-PS1-4** | Develop a model that predicts and describes changes in particle motion, temperature, and state of a pure substance when thermal energy is added or removed. |
| **MS-PS1-6** | Undertake a design project to construct, test, and modify a device that either releases or absorbs thermal energy by chemical process. |
| **MS-PS3-3** | Apply scientific principles to design, construct, and test a device that either minimizes or maximizes thermal energy transfer. |
| **MS-PS3-4** | Plan an investigation to determine the relationships among the energy transferred, the type of matter, the mass, and the change in the average kinetic energy of the particle as measured by the temperature of the sample. |

The Physical Sciences additions include topics on man-made materials and energy transfer, again reflecting current topics in science policy and allocation issues. Understanding not only how science works but how it is applied to society's current issues is an underlying goal of the NGSS. In addition to these applicable science standards, we also see the curriculum standards addressing the need to "model" and "experiment" with new ways and ideas to understand the issues that face modern society and how to best solve those challenges, as opposed to learning content knowledge alone (National Research Council, 2012).

*Why Are These Changes Important?*

The East Tennessee State University Center of Excellence in Mathematics and Science Education routinely makes use of a Project Management Team (PMT) to track the specific needs of our service area. The PMT membership consists of K-12 math and science teachers, K-12 principals and system-wide administrators, and higher education faculty in education and arts and sciences. Among needs reported most often by teachers and district administrators from targeted districts are the following: enhanced science content; embedded inquiry; an enhanced repertory of instructional strategies to meet student needs; and professional development aligned to the Tennessee science standards and to the forthcoming Next Generation Science Standards.

According to the 2009 Program for International Student Assessment (PISA), less than one third of fifteen-year-old U.S. students scored at the proficient level in science literacy, and the average U.S. science score lagged behind that of about 15 other countries (Fleischman et al., 2010). In the 2009 National Assessment of Educational Progress (NAEP), only about one third of fourth graders, about one third of eighth graders, and 21% of high school seniors were rated as proficient in science (Chapman, 2011). The NGSS framework addresses a deeper understanding and appreciation for science in our students and offers them the opportunities to perform authentic science investigations (Bell et al., 2012).

This article stems from a grant funded through the National Science Foundation Division of Graduate Education (Grant Number DGE-0742364; P.I. Dr. Gordon Anderson). This STEM Graduate Fellowship Program, entitled *Science First!*, is supported by East Tennessee State

University. The STEM fellows and K-5 teachers from North Side School of Math, Science, and Technology, a high need and racially/ethnically diverse elementary school, worked together for 6 years to develop and publish a collection of integrated lesson plans aligned to the Common Core State Standards and NGSS, using mathematics and science as a connecting thread. These guidemaps were created as a tool to be used by educators for curriculum development and revision and may be accessed on the GK-12 Program Website. (*Access website*: http://www.etsu.edu/cas/gk/).